\documentclass{intech}
\usepackage{multirow,multicol}
\usepackage{booktabs}

\chaptertitle{Hybrid Video Signal Coding Technologies: Past, Current and Future} 

\authors{Miaohui Wang, Ngan King Ngi \thanks{ The corresponding author is Miaohui Wang, Email: \texttt{wang.miaohui@gmail.com}; Ngan King Ngi is with the Department of Electronic
Engineering, The Chinese University of Hong Kong, Email: \texttt{knngan@ee.cuhk.edu.hk} }}
\affiliation{The Chinese University of Hong Kong} %
\country{Hong Kong}%

\begin{document}

\maketitle

\section{Introduction}
\label{Sec:introduction}
Video sequence generally contains a significant amount of statistical and perceptual redundancies. The ultimate goal of video coding is to reduce the average bits to represent the video signal by exploring these redundancies. Typically, the performance of video compression is related to both the amount of redundancy contained in the video data and the efficiency of the actual algorithms used for coding.

With the substantially increase in high-quality video service, it requires more advanced techniques, and hence many international video coding standards are developed in the past decades, including MPEG-2 \cite{tudor1995mpeg}, H.264/Advanced Video Coding (AVC) \cite{h26x_standard_h264_2003} and the latest High Efficiency Video Coding (HEVC) \cite{sullivan_hevc_overview_2012}. The standardization plays an important role in multimedia applications, which ensures interoperability and flexibility across a breadth of products made by different clients. To support communication across different platforms, the standards only define the syntax of the bitstream and describe the decoding method. Almost all the current standards are essentially block-based hybrid coding system, where video frame is split into non-overlapped blocks that are the basic coding unit. In a  modern video system, spatial correlation and spectral correlations can be removed by predictive and transform coding, whereas temporal correlation can be removed by motion compensated prediction.

As the cost for hardware has reduced and network support for coded video data has diversified, the need has arisen for video stream with higher resolution,  higher frame rate, or higher bit-depth.  If there are more pixels in each frame, more detailed information can be displayed.  Ultra high-definition (UHD) television supports 4K (3840$\times$2160) or 8K (7680$\times$4320) resolutions, which are becoming popular at increasingly low prices. In addition, increasing the number of frames displayed can  greatly improve the perceptual effect, especially for sport-castings and movies. At last, increasing the pixel bit width provides a more realistic scene, which can present the human eye with a comparable range of brightness and colors. Thus, how to efficiently compress the high-quality visual signal is the most fundamental driving force behind the development of next-generation video compression technology \cite{zeng2013perceptual}.

The primary of this chapter is summarized as follows. Section \ref{Sec:chapterFundmt} introduces the fundamentals of video coding platform. In Section \ref{Sec:STDS:past}, we  review various video compression technologies. The core parts of HEVC are introduced and discussed in Section \ref{Sec:STDS:hevc}. In Section \ref{Sec:STDS:future}, we present our developments for the next-generation video coding framework. In Section \ref{Sec:summary}, we summarize the contribution of the chapter embodied in this book, and suggest some the future research directions.

\section{Fundamentals in Video Coding Systems }
\label{Sec:chapterFundmt}
Video coding system is used to compress digital video signals to reduce the amounts of storage and transmission bandwidth. There are various types of video coding systems, such as the block-based, wavelet-based, and object-based systems. Nowadays, the block-based hybrid one is the most widely used and deployed in video compression. Examples of the block-based video coding systems include many international video coding standards such as the MPEG-1/2/4 part 2, H.264/MPEG-4 part 10 AVC \cite{h26x_standard_h264_2003} and the latest High Efficiency Video Coding (HEVC) \cite{sullivan_hevc_overview_2012} standards. This Section reviews the fundamentals of hybrid video coding systems. In this chapter,  YUV 4:2:0 video sequence with 8-bit or 10-bit depth will be used in the simulations, and Peak signal-to-noise ratio (PSNR) will be used as the main quality measure. 

\begin{figure*}[h!]
\centering
\begin{tabular}{c} 
 \includegraphics[width=0.95 \textwidth]{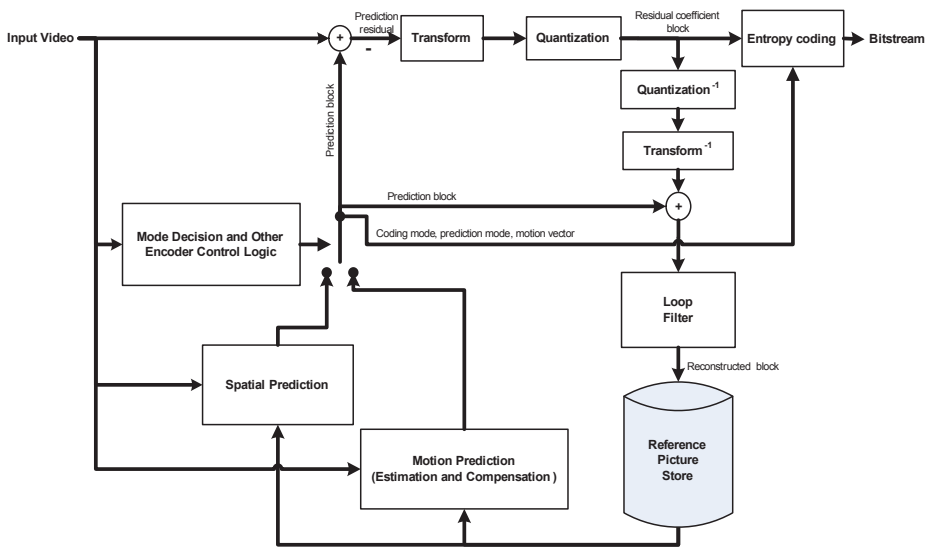}
\end{tabular} 
\caption{A general block diagram of a block-based video encoder.}  
\label{fig:Fundmt:intro_fig1}
\end{figure*}

Fig. \ref{fig:Fundmt:intro_fig1} gives the block diagram of a generic block-based hybrid video encoding system. The \textit{input video} signal is processed in a block-by-block manner. In the previous H.264/AVC standard, the basic block unit that commonly refers to as a macroblock (MB) consists of 16$\times$16 pixels \cite{rate_control_lagrangian_wang_2009}. Currently, JCT-VC (Joint Collaborative Team on Video Coding) of ITU-T/SG16/Q.6/VCEG and ISO/IEC/SC29/WG11/MPEG has developed the newest video coding standard called HEVC. In HEVC, the picture is coded by coding tree units (CTU) with equal size, such as 64$\times$64. The CTU is further divided into coding units (CU) with quad-tree partitioning. In HEVC, the CU size can be up to 64$\times$64 pixels, and it can be further partitioned into prediction units or PU, where separate prediction parameters are applied. For each input video block (MB or CU), \textit{spatial prediction} and/or \textit{temporal prediction}  may be performed. Spatial prediction (or ``intra prediction'') uses pixels from the already coded neighboring blocks in the same video picture/slice to predict the current video block. Spatial prediction reduces spatial redundancy inherent in the video signal. Temporal prediction (also referred to as ``inter prediction'' or ``motion compensated prediction'') uses pixels from the previously coded video pictures to predict the current video block, which is used to reduce temporal redundancy inherent in the video signal. Temporal prediction for a given video block is usually represented by one or more motion vectors that specify the amount and the direction of motion between the current block and its reference block. If multiple reference pictures are supported (as is the case for the recent video coding standards such as H.264/AVC or HEVC), then for each motion vector, its reference picture index is signaled additionally; and the reference index is used to identify the reference picture in the \textit{reference picture store}. After the spatial or temporal prediction, the \textit{mode decision block}  chooses the best prediction mode on the encoder side. The prediction block is then subtracted from the current \textit{input video} block; and the residual block is \textit{transformed}  and \textit{quantized}. The quantized coefficients are \textit{inverse quantized} and \textit{inverse transformed} to reconstruct the original residual, which is added back to the \textit{prediction block}  to form a reconstructed video block. Furthermore, \textit{loop filters}, such as deblocking filter (DF), Sample Adaptive Offset (SAO) and Adaptive Loop Filter (ALF), may be applied  on the reconstructed block before it is placed in the \textit{reference picture store} and referenced by future frames coding. To form the output video \textit{bitstream}, coding mode (inter or intra), prediction mode information, motion information, and quantized residual coefficients are all sent to the \textit{entropy coding unit} to be further compressed and packed in order.

\section{Video Signal Coding Technologies in Past H.264/AVC}
\label{Sec:STDS:past}
In past decades, video coding standards have been developed by two international organizations: Moving Picture Experts Group (MPEG) and  Video Coding Experts Group (VCEG). MPEG-x (x=1, 4) is recommended by MPEG, while H.26x (x=1, 3) is recommended by VCEG.  In addition, H.262/MPEG-2 \cite{tudor1995mpeg}, H.264/AVC \cite{h26x_standard_h264_2003} and HEVC \cite{sullivan_hevc_overview_2012} are jointly recommended by MPEG and VCEG. In this chapter, we first introduce three important standards, such as MPEG-2, MPEG-4 and H.264/AVC. Then,  we give a detailed description of the recent HEVC standard. The introductions for other video standards are referred to H.261,  MPEG-1 and H.263.  This section briefly reviews the state-of-the-art video coding standards, such as MPEG-2, MPEG-4, H.264/AVC and HEVC. HEVC is the latest international video standard, which adopts many new technical developments, including flexible block structure representation, residual quad-tree transform, sample adaptive offset as well as highly parallel processing architectures.

Nowadays, H.264/AVC is one of the most common international video coding standards, which is jointly developed by VCEG and MPEG. In contrast to MPEG-4 visual object (VO) coding,  H.264/AVC is of block-oriented video standard. In the past decade, H.264/AVC has been widely used for the recording, compression, and distribution of video signals. In this section, we highlight some of the features of H.264/AVC \cite{wang2012spatial}, including intra prediction, motion compensated prediction, transform, in-loop deblocking filter as well as entropy coding.

H.264/AVC bitstream contains four layers, such as GOP, frame, slice and block. The outermost layer is the video sequence, which consists of groups of pictures (GOP). Each GOP is made up of three kinds of frames, namely, I-, P- and B-frame. I-frame is the intra-coded frame, where only the spatial redundancy is reduced. It is used to prevent temporal error propagation. P-frame and B-frame are inter-coded using so-called reference frames, i.e., an I- or P-frame available in the reference store.

In the frame layer, an input frame is divided into non-overlapped macroblocks (MBs). Each MB consists of one 16$\times$16 Y-component, one 8$\times$8 U-component, and one 8$\times$8 V-component. The basic discrete cosine transform (DCT) block size is 4$\times$4.

\begin{figure*}[h!]
\centering
\begin{tabular}{c} 
 \includegraphics[width=0.6 \textwidth]{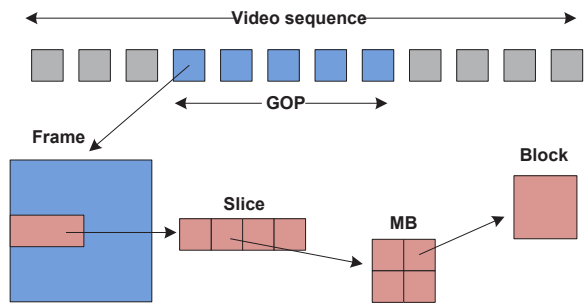}
\end{tabular} 
\caption{Layer structure used in H.264/AVC.}  
\label{fig:STDS:mpeg}
\end{figure*}

H.264/AVC employs a slice layer to maintain a constant output bit rate. Each slice contains several contiguous MBs in a raster scanning order, which can be encoded independently. The division of the frame into slices gives more flexibility to regulate the output bit rate. Fig. \ref{fig:STDS:mpeg} shows the four layers structure, including GOP, frame, slice and block levels.

\begin{itemize}
\item \textbf{Intra prediction}. H.264/AVC consists of two different types of intra prediction modes for luma channel, such as Intra$\_$4$\times$4 and Intra$\_$16$\times$16. In the Intra$\_$4$\times$4 mode, a macroblock is divided into no-overlapping 4$\times$4 blocks, and the prediction is applied to each 4$\times$4 individually.  The Intra$\_$4$\times$4 mode supports nine directional predictions. 

In the Intra$\_$16$\times$16 mode, only one prediction mode is applied for the whole macroblock. The Intra$\_$16$\times$16 mode supports four different directional predictions: vertical prediction, horizontal prediction, DC-prediction and plane-prediction. The prediction operations of 16$\times$16 modes are the same as that of 4$\times$4 prediction modes. All the intra prediction modes are explained in detail in \cite{richardson2004h}.

\item \textbf{Motion compensated prediction}. In comparison to prior video-coding standards,  H.264/AVC supports variable block-size motion compensation, including 16$\times$16, 16$\times$8, 8$\times$16, 8$\times$8, 8$\times$4, 4$\times$8, and 4$\times$4. For each  8$\times$8 or larger block, the motion vectors can point to different reference frames. In order to estimate and compensate the half-pixel luma sample prediction, the reference frame is interpolated by a 6-tap finite impulse response (FIR) filter.  By averaging the samples at half- and integer-pixel positions,  the sample at quarter-pixel positions is generated. In addition,  weighted prediction is allowed in performing motion compensation. Details on  motion-compensated prediction can be found in \cite{richardson2004h}.

\item \textbf{Transform coding}. In the former standards such as MPEG-1 and MPEG-2, a 2D DCT is applied to the 8$\times$8 block.  Instead of using DCT, H.264/AVC reference software supports two different integer transforms, including 4$\times$4  and 8$\times$8. The new integer transform is conceptually similar to DCT but is able to provide exactly specified decoding.

Besides using the integer transform in the luma component, H.264/AVC adopts Hadamard transform to further improve the coding performance. In the Intra$\_$16$\times$16 mode,  the macroblock is first transformed by  the 4$\times$4 integer transform. There are total 16 DC coefficients in an Intra$\_$16$\times$16 coded macroblock. To further reduce the correlation between these DC coefficients, a 4$\times$4 Hadamard transform is applied as the second transform. 

The transform coefficients are quantized by a scalar quantizer in which the quantization step size is chosen by a quantization parameter.  The quantization step size doubles when the quantization parameter value is increased by 6.  In H.264/AVC, there are 52 different quantization parameters. The transform is explained in detail in \cite{malvar2003low}. To improve the coding performance, directional transforms \cite{xu2012video} and content adaptive transforms \cite{ wang2013efficient, wang2013rate, cat_wang2014} have been widely studied.

\item \textbf{In-loop deblocking filter}. In the block-based video coding, the reconstructed frame usually suffers from visually annoying effects known as the blocking artifacts. In  H.264/AVC, the block artifacts can be caused by two coding tools. One is the transform coding, and the other is the motion compensated  prediction. In transform coding, the boundary pixels of adjacent blocks can easily become discontinuous due to the block-based quantization. In motion compensated prediction, the prediction block is generated by copying the interpolated pixels from different locations of possibly different reference frames, which causes discontinuities on the edges of copied blocks.

H.264/AVC employs an efficient in-loop filter after the inverse transform in the encoder and decoder, respectively. The filter is applied to each macroblock to reduce the blocking artifacts, which is applied to the vertical or horizontal block edges except for the edges on the slice boundaries. A detailed description of the H.264/AVC deblocking can be found in \cite{huang2003architecture}.

\item \textbf{Entropy coding}. H.264/AVC specifies two alternative entropy coding methods: a low-complexity technique of context-adaptive variable length coding (CAVLC), and a high-efficient algorithm of context-based adaptive binary arithmetic coding (CABAC). By incorporating context modeling in the entropy coding framework,  H.264/AVC offers a high degree of adaptation to the underlying data source.
\end{itemize}

\section{Video Signal Coding Technologies in Present HEVC}
\label{Sec:STDS:hevc}
HEVC is the latest international standard for high-resolution video signals, which contains the most advanced video coding methods. As compared to its predecessors, HEVC is designed to provide higher coding efficiency, higher resolution and more sophisticated multimedia applications. HEVC  belongs to the traditional block-based hybrid video coding system, but it employs a large number of new technologies, including highly flexible block structure representation, residual quad-tree transform (RQT), sample adaptive offset (SAO) as well as highly parallel processing architectures.

\subsection{Coding Block Structures}
\label{blk_structure}
In HEVC, the input frame is divided into coding tree units (CTUs) as large as 64$\times$64 luma samples \cite{wang2015lowdelay}, which can be is broadly considered as analogous to macroblocks in H.264/AVC \cite{yan2009adaptive}.  CTU represents the basic processing unit, which consists of  coding units (CUs) that is a square region.  The CUs inside a CTU are coded in a z-scan order. For each CU, one of the prediction modes  (i.e., intra, inter or skip) is signaled in the bitstream.  In the non-skip mode,  the current CU  is coded by the regular process with a prediction mode, either intra or inter prediction. In the skip mode, the current CU is considered to be inter-coded  without encoding of motion vector differences (MVD) and residual information.

To enhance the coding efficiency, HEVC enables the CU to be divided into one, two or four prediction units (PUs) based on the partition mode. PU defines a region having the same prediction information. In HEVC, PU can take a size ranging from 64$\times$64 to 4$\times$4. It should be noted that the minimum PU size is 8$\times$8 in an inter-coded PU to save the memory bandwidth.  Intra-coded CU only supports PART$\_$2Nx2N and PART$\_$NxN, whereas inter-coded CU supports all partition modes.

The transform unit (TU) is the basic unit for transform coding, which is a square region sharing the same transform and quantization. By supporting various transform block sizes, the residual quad-tree transform (RQT) enables the adaptation of transform to the varying space frequency characteristics of the residual signal. In HEVC, the size of TU can be ranging from 32$\times$32 to 4$\times$4. In an inter-coded CU, the size of TU can be larger than that of PU, while the size of TU must not be larger than that of PU in an intra-coded CU.

\subsection{Transform and quantization}
In video coding, transforms are applied to the residual block resulting from inter or intra  prediction. In HEVC,  the transform matrices are an approximation of the traditional DCT matrices. For intra 4$\times$4 TU, an integer approximation of discrete sine transform (DST) is applied to the luma component. In the current \cite{hevc_software}, the transform is performed by partial butterfly structure for low computational complexity. 

Let $C$ be the traditional DCT matrix. The DCT approximation used in HEVC is $C = {r_N}{C_N},{r_N} = {\raise0.5ex\hbox{$\scriptstyle 1$}
\kern-0.1em/\kern-0.15em
\lower0.25ex\hbox{$\scriptstyle {(64\sqrt N) }$}}$, where $N$ is the block size. Then, the forward transform of HEVC is,
\begin{equation}
\begin{array}{ll}
\bar{Y} & = \left( {{r_N}{C_N}} \right) \times X \times {\left( {{r_N}{C_N}} \right)^T}\\
 &= r_N^2\left( {{C_N} \times X \times {C_N}^T} \right)\\
 &= {2^{ - 12 - {{\log }_2}N}}\left( {{C_N} \times X \times {C_N}^T} \right),
\end{array}
\label{eq:hevc:forward_hevc_transform}
\end{equation}
where $\bar{Y}$ and $X$ are the transform coefficients and the input residual block, respectively. In order to maintain the output results within 16-bits, HEVC takes two stages of forward transform. Thus, equation (\ref{eq:hevc:forward_hevc_transform}) is re-written as: 
\begin{equation}
\begin{array}{ll}
\bar{Y} &= {2^{ - 7 + {{\log }_2}N}}\left\{ {{2^{ - 6 - {{\log }_2}N}}\left[ {{2^{1 - {{\log }_2}N}}{C_N} \times X} \right] \times {C_N}^T} \right\}\\
 &= {2^{ - 7 + {{\log }_2}N}} \hat Y,  
\end{array}
\end{equation}
where $\hat Y$ is the output of the HEVC forward transform. It should be noted that the results of the first stage of the forward transform is right-shifted by ${\log}_2 N -1$, while the results of the second stage of forward transform is right-shifted by ${\log}_2 N +6$. ${2^{ - 7 + {{\log }_2}N}}$. In addition, the right-shifted operation is integrated into the quantization process in the reference software of HEVC.

The quantization process in HEVC is described as follows. 
\begin{equation}
\begin{array}{ll}
{\bar{Y}_q}  &= \frac{{\overline Y }}{{{Q_{step}}}}\\
& = \frac{{\hat Y  }}{{{Q_{step}}}}\frac{1}{{{2^{7 - {{\log }_2}N}}}}\\
 &= \frac{{\hat Y }}{{{Q_{step}}}}\frac{1}{{{2^{7 - {{\log }_2}N}}}}
 \label{eq:hevc:forward_quantization}
\end{array}
\end{equation}

In the HEVC reference code HM16.0, the relationship between $Q_{step}$ and $QP$ is $f\left( {QP\% 6} \right) = \frac{{{2^{{\raise0.5ex\hbox{$\scriptstyle {QP}$}
\kern-0.1em/\kern-0.15em
\lower0.25ex\hbox{$\scriptstyle 6$}}}} \times {2^{14}}}}{{{Q_{step}}\left( {QP\% 6} \right)}}$, where $f \left( {QP\% 6} \right) = \{26214, 23302, 20560, 18396, 16384, 14564 \}$.  Equation (\ref{eq:hevc:forward_quantization}) is then rewritten as ${\bar{Y}_q} = \hat Y \times f\left( {QP\% 6} \right) \times \frac{1}{{{2^{21 - {{\log }_2}N + }}{\raise0.5ex\hbox{$\scriptstyle {QP}$}
\kern-0.1em/\kern-0.15em
\lower0.25ex\hbox{$\scriptstyle 6$}}}}$,  where  ${{2^{21 - {{\log }_2}N + }}{\raise0.5ex\hbox{$\scriptstyle {QP}$}
\kern-0.1em/\kern-0.15em
\lower0.25ex\hbox{$\scriptstyle 6$}}}$ is termed as \textit{iQBits} in  the reference software of HEVC.

The de-quantization is performed as
\begin{equation}
\begin{array}{ll}
{\bar{Y} _{{q^{ - 1}}}} &= {\bar Y _q} \times {Q_{step}}\\
 &= {\hat Y _q} \times {f^{ - 1}}\left( {QP\% 6} \right) \times {2^{{\raise0.5ex\hbox{$\scriptstyle {QP}$}
\kern-0.1em/\kern-0.15em
\lower0.25ex\hbox{$\scriptstyle 6$}} - 6}}, 
\end{array}
\end{equation}
where  ${f^{ - 1}}\left( {QP\% 6} \right)=\{ 40, 45, 51, 57, 64, 72\}$

The inverse transform is given in equation (\ref{eq:hevc:backward_hevc_transform}). Similarly, all the intermediate results need to maintain 16-bits precision. Thus, the results of the first stage of inverse transform  will be right-shifted by 7, while the results of the second stage of inverse transform  will be right-shifted by 12. 
\begin{equation}
\begin{array}{ll}
\widetilde{Y}'  &= {\left( {{r_N}{C_N}} \right)^T} \times {\bar Y _{{q^{ - 1}}}} \times \left( {{r_N}{C_N}} \right)\\
 &= r_N^2\left( {{C_N}^T \times {{\bar Y }_{{q^{ - 1}}}} \times {C_N}} \right)\\
 &= {2^{ - 12 - {{\log }_2}N}}\left( {{C_N}^T \times {{\bar Y }_{{q^{ - 1}}}} \times {C_N}} \right)\\
 &= {2^{ - 7 + {{\log }_2}N}}\left\{ {{2^{ - 12}}\left[ {{2^{ - 7}}{C_N}^T \times {{\bar Y }_{{q^{ - 1}}}}} \right] \times {C_N}} \right\}\\
 &= {2^{ - 7 + {{\log }_2}N }} \times  \widetilde{Y} 
\end{array}
\label{eq:hevc:backward_hevc_transform}
\end{equation}
where $\tilde{Y}$ is the reconstructed residual block.  It is noted that $ {2^{ - 7 + {{\log }_2}N }}$ is integrated into the inverse quantization process. An improved quantization method is introduced in \cite{wang2015quantization}.

\subsection{In-loop Filters}
HEVC adopts two processing stages in the in-loop filter unit. In the first stage, a deblocking filter (DF) is used to reduce the visibility of blocking artifacts, which is applied only to samples located at the block boundaries. In the second stage, a sample adaptive offset (SAO) filter is used to improve the accuracy of the reconstruction of the original signal amplitudes, which is applied adaptively to CTU by CTU.

In DF, the vertical boundaries are filtered first, followed by the horizontal boundaries. For the vertical boundaries, the filtering order is from the left-most boundary to the right-most one. For the horizontal boundaries, the filtering order is from the top-most boundary to the bottom one.  Boundary strength (BS) is used to indicate the DF decision, where higher value of \textit{BS} means a stronger filtering effect. Specifically, the filter is only applied to the block boundaries with \textit{BS} greater than zero for a luma component, whereas the filter is applied to chroma components only if \textit{BS} is greater than one. The detailed description can be found in \cite{norkin2012hevc}.

In SAO, edge offset (EO) and band offset (BO) are used to attenuate ringing artifacts which are more likely to appear if larger transform is used.  For EO, the sample classification is based on the comparison between the current samples and neighboring samples in some direction. There are total four edge directions: horizontal, vertical, 45 degree diagonal and 135 degree diagonal. If the EO mode is chosen, one of the best edge patterns will be selected, and the associated pattern index and the absolute offset are encoded in the bitstream. Simulation results show that EO can be used to reduce undesired sharp edges \cite{fu2012sample}, and hence more pleasing details can be restored. For BO, the input block is equally divided into several bands according to the sample bit depth. If the input sample is 8-bit depth, data will be split into 32 bands, where the width of each band is 8. When we say one sample belongs to band $k$, the value of this sample must be in $8k$ to $8k+7$ inclusive, where $k$ ranges from 0 to 31.  For each band, the average difference (or called ``offset'') between the original and reconstructed samples is obtained. Then, only offsets and the initial band position of those four consecutive bands will be selected in terms of rate distortion optimization, and the related side information is encoded in the bitstream. It is noted that the signs of offsets in BO need to be encoded, which is different from that of in EO.

\section{Video Signal Coding Technologies in Future }
\label{Sec:STDS:future}
Interest in deploying new services, such as high resolution, high frame rate and high bit depth video signals, has been driven by the decreasing cost of transmission and storage bandwidth. Typically, 720 progressive (720p) or 1080 progressive (1080p) or 1080 interlace (1080i) video resolutions are known as high-definition (HD), while Ultra HD includes 4K (3840$\times$2160) or 8K (7680$\times$4320) resolutions. Additionally, high frame can help greatly improve the human perceptual effect, especially for the sportcastings and movies. Furthermore, high bit depth video provides a more real-world scene, which can present the human eye with a comparable range of colors. Therefore, how to efficiently compress high-quality video signals is essential in the development of next-generation video coding.

This section focuses on improving the coding block structure of the input video signal, adaptive loop filter and sample adaptive offset for next-generation video coding. In particular, the basic video block size for processing is extended to super-block or super coding unit (SCU) in Section \ref{sec:NextG:superblock}, including Direct-CTU and SCU-to-CTU modes. Additionally, the adaptive loop filter  and  sample adaptive offset methods are also investigated based on super-block encoding structure in Sections \ref{sec:NextG:cu_level_alf} and \ref{sec:NextG:adaptive_sao}, respectively.  In the rest of this chapter, the terms ``super-block'' and ``SCU'' are used interchangeably.

\subsection{Super-block Encoding Structure}
\label{sec:NextG:superblock}
The current design of the HEVC is based on the quadtree structure \cite{wang2015efficient}. High-resolution video coding benefits from a larger prediction block size and thereof transform and quantization of prediction residues. However, in the current HEVC video coding standard, the maximum CTB size is 64$\times$64, which can limit a possible larger prediction block in Ultra HD video coding, and hence cause negative effects on coding efficiency. It is possible to extend CTB to a super-block with a larger block size for the Ultra HD video coding. Super-block signaling for homogeneous area can save the overhead signaling of the mode of coding unit, the PU partition type and the prediction related information (e.g., intra prediction mode, motion vector) compared to the existing HEVC design. However, the encoding complexity will be increased significantly, which is the major bottle-neck for the super-block based video coding system.  In order to adopt super-block structure into the current HEVC scheme, we need to take into account both the coding efficiency and encoding complexity. Specifically, we propose to process a super-block using two encoding modes: Direct-CTU and SCU-to-CTU; Direct-CTU is designed for the complex region while SCU-to-CTU for the homogeneous region in a picture; the best coding setting applied to a super-block is chosen from these two modes.

To solve the problem of applying super-block or super coding unit (SCU) for Ultra HD video coding as mentioned above we propose to use two separate coding structures to encode a super-block, including Direct-CTU and SCU-to-CTU modes. In Direct-CTU, an SCU is first split into a number of predefined CTUs, and then, the best encoding parameters are searched from the current CTU to the possible minimum coding unit (MCU). Similarly, in SCU-to-CTU, the best encoding parameters are searched from SCU to CTU. It is noted that the size of CTU is configurable, and one typical size in our method is 64$\times$64. The best coding mode for a super-block is chosen from these two modes based on the overall rate-distortion cost of the super-block.

Next we discuss the benefits of the proposed scheme, 1) Direct-CTU mode, and 2) SCU-to-CTU mode. Note that the associated syntax changes can be found from my PhD thesis that is given in Section \ref{Sec:summary}. 

\begin{figure*}[h!]
\centering
\begin{tabular}{c} 
 \includegraphics[width=0.6 \textwidth]{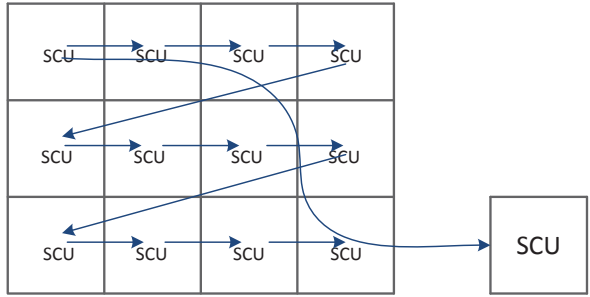}
\end{tabular} 
\caption{Partitioning with super-blocks in a frame.}  
\label{fig:NextG:background_fig7}
\end{figure*}

\subsubsection{Direct-CTU Mode}
\label{sec:NextG:superblock:proposed_method:direct_ctu_mode}
In our method, an input video frame is first divided into non-overlapping SCUs.  SCU is a square block with equal size, and is encoded in the raster scanning order within one frame. As shown in Fig. \ref{fig:NextG:background_fig7}, there are totally 12 SCUs in one frame. For each SCU, both Direct-CTU and SCU-to-CTU methods are conducted to find the best encoding parameters (i.e., coding mode, prediction mode, motion vector, quantized  coefficients, etc.).

In Direct-CTU mode, the input SCU is directly divided into many CTUs with the equal size. Let $M_{SCU}$, $M_{CTU}$  and $M_{MCU}$ denote the size of SCU, CTU and MCU, respectively.  There are four configurable parameters that are used to represent the relationships among them:

\begin{itemize}
\item \textbf{MaxSCUWidth} specifies the maximum width of a super-block.
\item	\textbf{MaxSCUHeight} specifies the maximum height of a super-block.
\item	\textbf{MaxPartitionDepth} specifies the depth of MCU relative to SCU in quadtree structure.
\item	\textbf{MaxDirectPartitionDepth} specifies the depth of the CTU relative to SCU in quadtree structure.
\end{itemize}

In the proposed method, $M_{CTU}$ = $2^{{log}_2 \left( M_{SCU} -\textbf{MaxDirectPartitionDepth} \right)}$; $M_{SCU}$=\textbf{MaxSCUWidth} or \textbf{MaxSCUHeight}; $M_{MCU}$ = $2^{ {log}_2 \left( M_{SCU} -\textbf{MaxPartitionDepth} \right)}$. Both \textbf{MaxPartitionDepth} and \textbf{MaxDirectPartitionDepth} are set as non-negative integer values including 0.  \textbf{MaxPartitionDepth} being equal to 0 indicates that there is no partition for a SCU. Additionally, \textbf{MaxDirectPartitionDepth} must be no greater than \textbf{MaxPartitionDepth}, which means the size of CTU is always no smaller than the size of MCU. When they are configured with the same value, the Direct-CTU mode is degraded as the SCU-to-CTU mode. In this case, only the Direct-CTU method would be processed while SCU-to-CTU is bypassed.

\begin{figure*}[h!]
\centering
\begin{tabular}{cc} 
 \includegraphics[width=0.35 \textwidth]{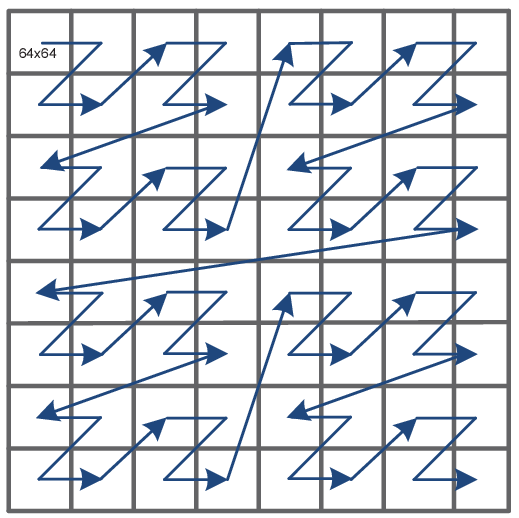}
&\includegraphics[width=0.35 \textwidth]{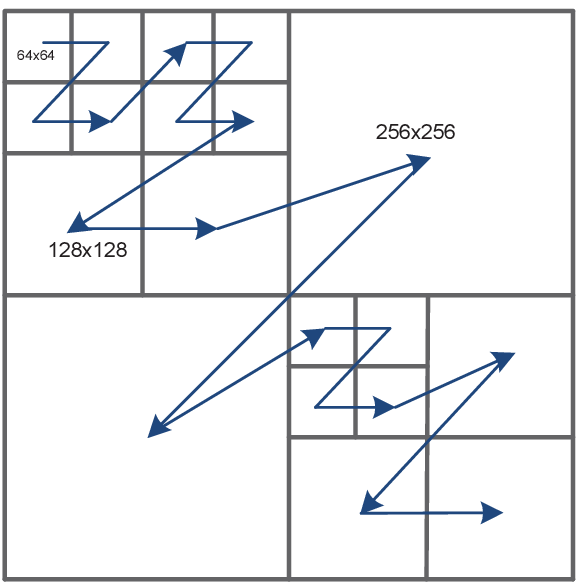}
 \\ 
 (a)&(b)\\
\end{tabular} 
\caption{Direct-CTU mode and SCU-to-CTU modes.}  
\label{fig:NextG:background_fig8}
\end{figure*}

The encoding order in Direct-CTU mode is illustrated in Fig. \ref{fig:NextG:background_fig8} (a). The super-block size is 512$\times$512 ($M_{SCU}$ = 512), and the size of CTU is 64$\times$64 ($M_{CTU}$ = 64). We propose to encode CTUs using the Z-scanning order in a super-block, where it belongs to the depth-first search (DFS).  Compared to the raster scanning order, DFS contributes to a higher coding efficiency. The reason is that both the left and the top neighboring CTUs have been encoded/transmitted before the current block, and the CABAC context is more efficiently in the Z order than in the raster order. Consequently, the encoded neighboring CTUs can be used to facilitate the encoding of the current block (e.g., for motion vector prediction or context modeling in entropy coding). In Direct-CTU mode, each CTU is encoded by the current HEVC scheme, and the possible minimum block size is $M_{MCU}$ = $2^{ {log}_2 \left( M_{SCU} -\textbf{MaxPartitionDepth} \right)}$.

\subsubsection{SCU-to-CTU Mode}
\label{sec:NextG:superblock:proposed_method:scu_to_ctu_mode}
In SCU-to-CTU mode, the SCU is divided into CUs with variable sizes by the recursive quadtree partition.  In one design of the proposed method, the minimum possible CU size in SCU-to-CTU mode is $M_{CTU}$, where \textbf{MaxDirectPartitionDepth} can be considered as an early termination mechanism set up beforehand. The reason is that as the size of SCU increases, the encoding for testing all possible CU partitions (e.g., from SCU to MCU) becomes too complex.  Thus, the computational complexity of employing super-blocks in Ultra HD videos can be expected as one of the most challenging tasks for next-generation video coding.  On the other hand, Direct-CTU is designed for the complex region in which small CUs are commonly selected for a higher video quality, while SCU-to-CTU is designed for the homogeneous region where large CUs are commonly selected for a better coding efficiency. The combination of Direct-CTU and SCU-to-CTU modes can greatly facilitate the deployment of super-block for Ultra HD video coding.

 A typical example of the SCU-to-CTU process is illustrated in Fig. \ref{fig:NextG:background_fig8}  (b).  The current super-block is 512$\times$512 ($M_{SCU}$ = 512), and the CTU size is 64$\times$64 ($M_{CTU}$ = 64). In SCU-to-CTU mode, each CU is encoded by the current HEVC scheme.

\begin{figure*}[h!]
\centering
\begin{tabular}{cc} 
 \includegraphics[width=0.35 \textwidth]{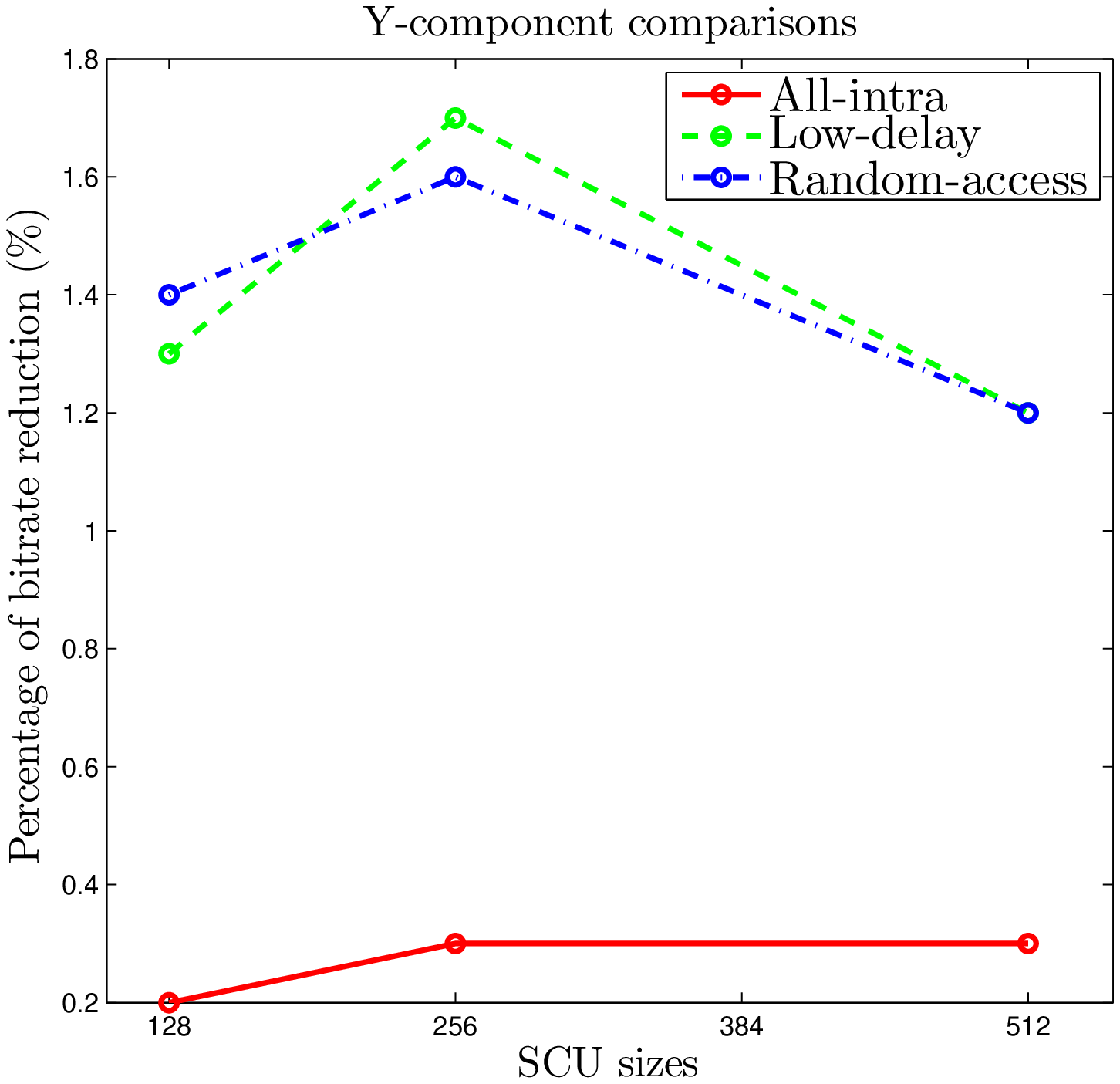}
&\includegraphics[width=0.35 \textwidth]{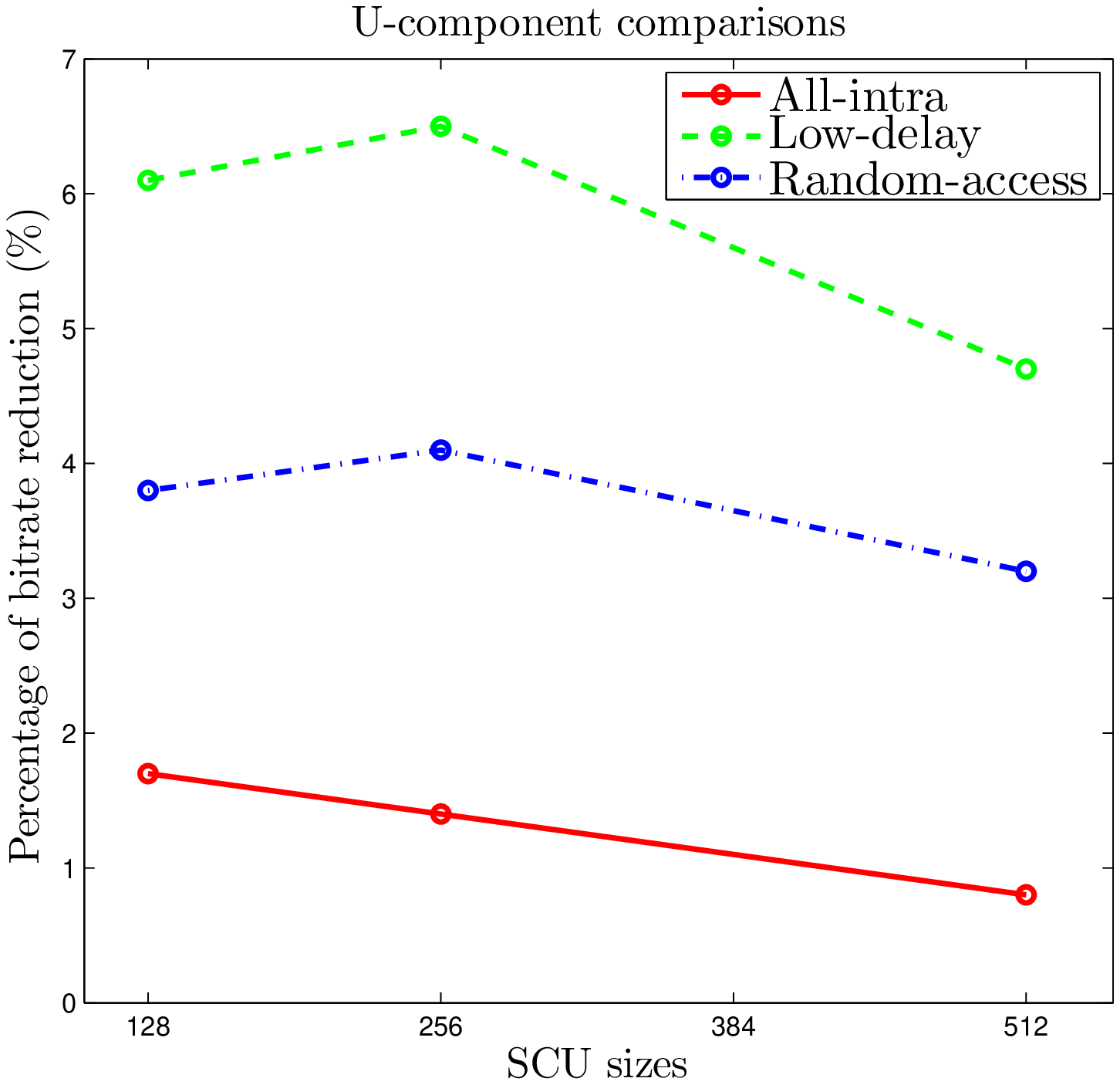}\\
 (a)&(b)\\
\includegraphics[width=0.35 \textwidth]{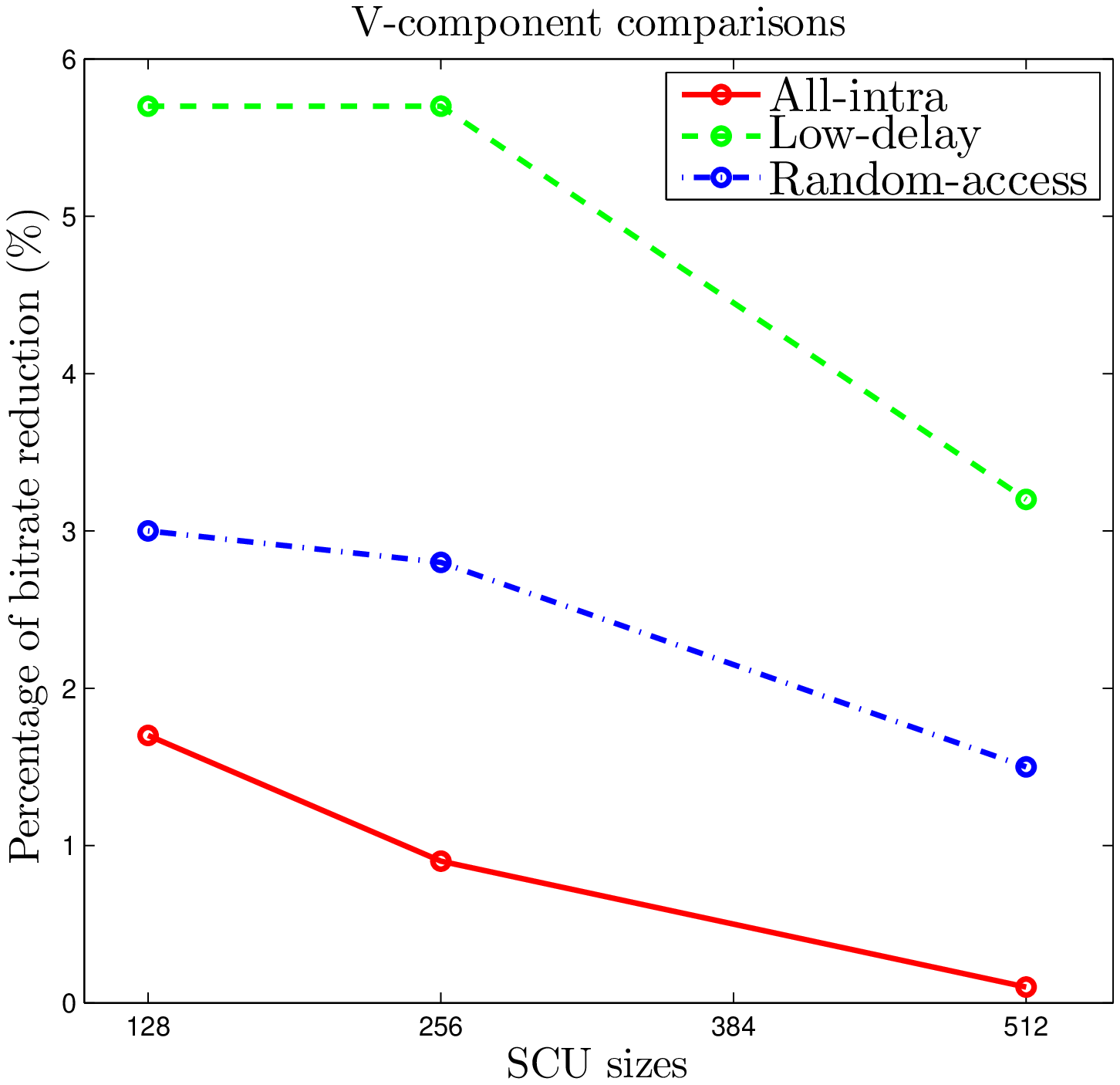}
 &\\ 
(c) &\\
\end{tabular} 
\caption{BD-bitrate comparisons under different SCU sizes.}
\label{fig:NextG:scu_rd_curves}
\end{figure*}

\subsubsection{Simulation Results}
\label{sec:NextG:superblock:simulation_results}
In this section, we evaluate the performance of the proposed method in comparison with HM16.0.  In the experiments, there are total 11 multiple-resolutions test sequences (i.e., tabulated in Table \ref{tab:NextG:sao:result_Y_ld}), including HD and  ultra-HD formats. Also, 10-bit ultra-HD video sequences are employed to validate its performance.

To find the best super-block size for Ultra-HD videos, we compare our method with HM16.0 under another three super-block sizes: SCU=128$\times$128, 256$\times$256 and 512$\times$512. Fig. \ref{fig:NextG:scu_rd_curves} shows the comparison results.  One can see that the coding performance of SCU=256$\times$256 is slightly better or comparable performance in comparison with that of SCU=128$\times$128 and SCU=512$\times$512. Additionally, we also compare the average encoding and decoding complexities under three configuration profiles. For example, the average encoding  complexities of our method are 180$\%$, 250$\%$ and  254$\%$ for SCU=128$\times$128, SCU=256$\times$256 and 512$\times$512, respectively, in comparison with HM16.0.  The average decoding  complexities of our method are 94$\%$, 89$\%$ and  89$\%$ for SCU=128$\times$128, SCU=256$\times$256 and 512$\times$512 , respectively, compared to HM16.0.  It is observed that the encoding and decoding complexities of SCU=256$\times$256 and SCU=512$\times$512 are similar to each other, which can be explained by the percentage of CU sizes. In the experiments with setting SCU=512$\times$512, the average usage of CU=512$\times$512 is less than 1$\%$. At the same time, since SCU=512$\times$512 increases the coding depth, it needs more bits to signal the related partitions, which indicates that SCU=512$\times$512 is less efficiency than other SCU sizes, such as 128$\times$128 and 256$\times$256.

\subsection{Improved Coding Unit Level Adaptive Loop Filter}
\label{sec:NextG:cu_level_alf}
\begin{figure*}[h!]
\centering
\begin{tabular}{c} 
 \includegraphics[width=0.8 \textwidth]{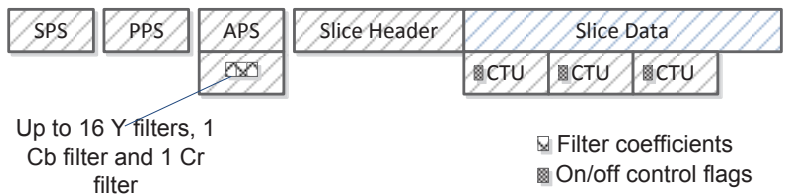}
\end{tabular} 
\caption{Basic syntax structure of ALF.}  
\label{fig:NextG:alf:background_fig6}
\end{figure*}

As the CTB is extended to super-block, the basic block unit processed by ALF is also directly extended to the same size according to the existing syntax structures of ALF \cite{tsai2013adaptive} and the related syntax design is depicted in Fig. \ref{fig:NextG:alf:background_fig6}. However, a larger block unit will cause negative effect on the original purpose of region-based filter adaptation in ALF. Since the size of super-block is too large, the determination of on/off control is compromised. Furthermore, it may cause over-smoothing.  To avoid this problem, CU level filter on/off control will be considered. Specifically, we propose to apply CU level based ALF to filter a super-block, where three types of on/off control flag are included: slice level flag, super-block level flag, and CU level flag. If both the slice level and super-block level flag are on, the CU level filtering will be employed and the associated flag will be signaled to the decoder.

When the slice level flag, super-block level flag and all CU level flags within that super-block are simultaneously on, it indicates that all CUs in the current super-block are needed to be filtered. Such a case may result in a large number of overhead bits.  Hence, we propose an improved CU level based ALF method, where an additional super-block level flag is used to indicate whether all CU flags in the current super-block are on or not. If the additional super-block level flag is on, no CU level flag will be signaled, and the decoder can infer that ALF is applied for all CUs within the super-block. If this additional flag is off, the CU level on/off flags will be explicitly encoded to indicate to the decoder whether ALF filtering is applied for each CU.

\subsubsection{Proposed CU-level ALF method}
\label{sec:NextG:cu_level_alf:proposed_method}
Extensive simulation results show that the coding performance of ALF is significantly improved when the input video resolution increases. From the coding efficiency perspective, ALF can be helpful for video compression in large resolutions. On the other hand, as a super-block coding structure is applied, the CTB-level ALF, in HM16.0, is directly extended to super-block level ALF. However, it may result in over-smoothing that is one of the adverse effects of repetitive use of Wiener filter (i.e., ALF).

To tackle the problem in super-block level ALF, we propose to change the filtering control from super-block level to the CU level, which leads to a more accurate control with a finer granularity. In addition, an improved CU level ALF signaling method in Section \ref{sec:NextG:cu_level_alf:proposed_method:improved_cu_alf}  is also proposed to further improve the coding efficiency.  In the next sections, we discuss the benefits of the proposed scheme of an improved CU-level ALF method, together with a few associated syntax changes \cite{tsai2013adaptive}. 

\begin{figure*}[h!]
\centering
\begin{tabular}{c} 
 \includegraphics[width=0.8 \textwidth]{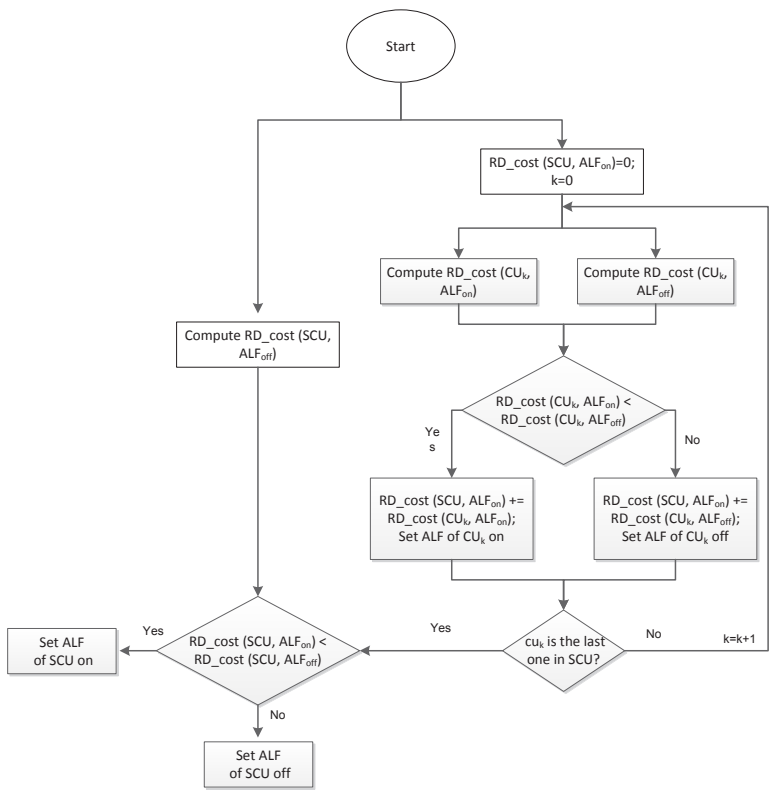}
\end{tabular} 
\caption{Flowchart of CU level flag determination. RD$\_$cost (argument 1, argument 2) is used to record the rate distortion cost of the input encoding modes, where the argument 1 is the block type and the argument 2 is the status of ALF flag.}  
\label{fig:NextG:alf:flowchart_fig9}
\end{figure*}

\subsubsection{CU-level ALF}
\label{sec:NextG:cu_level_alf:proposed_method:cu_alf}

When CTU is extended to super-block, the CTU level ALF is also extended to the same size according to the syntax structures of ALF in \cite{tsai2013adaptive}. As mentioned previously, applying ALF at super-block level may cause over-smoothing. Block level adaptive ALF control flag \cite{tsai2013adaptive} is a better way to avoid the adverse effects of Wiener filter on a super-block. Specifically, a CU-level ALF on/off control flag is added. If the rate-distortion cost with ALF applied is less than that without ALF applied in the current CU, the control flag is set to 1 to indicate that ALF is applied. Otherwise, the CU-level ALF flag is set to 0 to indicate that ALF is not applied. The benefits of CU-level ALF are summarized as: (1) the inaccurate filtering problem on super-block can be improved, and (2) the latency of the decoder can also be improved because CUs with the ALF off flags in the current super-block will not be filtered.

The CU-level ALF control flag is determined as shown in Fig. \ref{fig:NextG:alf:flowchart_fig9}: Firstly, the rate-distortion costs of ALF on and off for each CU are separately computed, the minimum cost between them is considered as the cost for the current CU, and the summation cost of all CUs in the super-block is obtained. Secondly, the rate-distortion cost with super-block level ALF off is also computed. Finally, the minimum cost is selected between these two modes, and the associated mode information will be coded. If the first mode is chosen, the CU-level flags are coded as side information.

\begin{figure*}[h!]
\centering
\begin{tabular}{c} 
 \includegraphics[width=0.35 \textwidth]{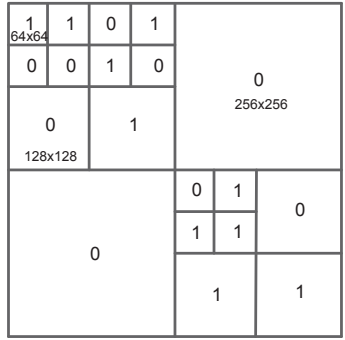}
\end{tabular} 
\caption{CU-level flagging.}  
\label{fig:NextG:alf:flowchart_fig10}
\end{figure*}

\subsubsection{Improved CU-level ALF}
\label{sec:NextG:cu_level_alf:proposed_method:improved_cu_alf}

Fig. \ref{fig:NextG:alf:flowchart_fig10} shows a typical example of CU-level ALF flagging. In this super-block, the maximum filtered CU is 256$\times$256, and the minimum filtered CU size is 64$\times$64. CUs with control flags equal to 1 will be filtered by ALF, while flags equal to 0 will not be filtered. It can be seen that if the traditional ALF method (i.e., super-block level based ALF) is used, the current block size filtered by ALF is 512$\times$512. It will amplify the adverse effect of Wiener filter and causes a longer latency.

The above CU level ALF method in Section \ref{sec:NextG:cu_level_alf:proposed_method:cu_alf} may suffer from the overhead problem. For example, there is an extreme case: both the super-block-level and all CU-level flags are 1. If the SCU is split in CUs, it costs a lot of bits to signal the flags all equals to 1. To avoid this situation, we propose to add a new additional super-block level flag to indicate if all CUs are filtered. The detailed relationships among ALF flags are tabulated in Table \ref{tab:alf:syntax_2}.

\begin{table}[!h]
\centering
\caption{Coding tree unit syntax in HEVC} 
\label{tab:alf:syntax_2}
\newsavebox{\tablebox}
\begin{lrbox}{\tablebox}
\begin{tabular}{|l|c|c|}
\hline
super-block                 &\multicolumn{2}{|c|}{Improved CU-level ALF flagging}\\
\cline{2-3}
level flagging				&All CU filtered flagging	&CU level flag (0: off, 1:on)\\
\hline
0							& \textit{inferred as 0}				& \textit{inferred as 0}\\
\hline
1							&1							& \textit{inferred as 1}\\
\hline
							&0							&signaled \\
\hline
\end{tabular}
\end{lrbox}
\scalebox{0.90}{\usebox{\tablebox}}
\end{table}

Table \ref{tab:alf:syntax_2} shows that if the additional super-block level flag is 1, there is no CU level flag coded and the decoder will derive and set all CU-level flags equal to 1.  If the additional super-block level flag is 0, only in this case, the CU-level flags will be coded as side information.

\begin{figure*}[h!]
\centering
\begin{tabular}{cc} 
 \includegraphics[width=0.40 \textwidth]{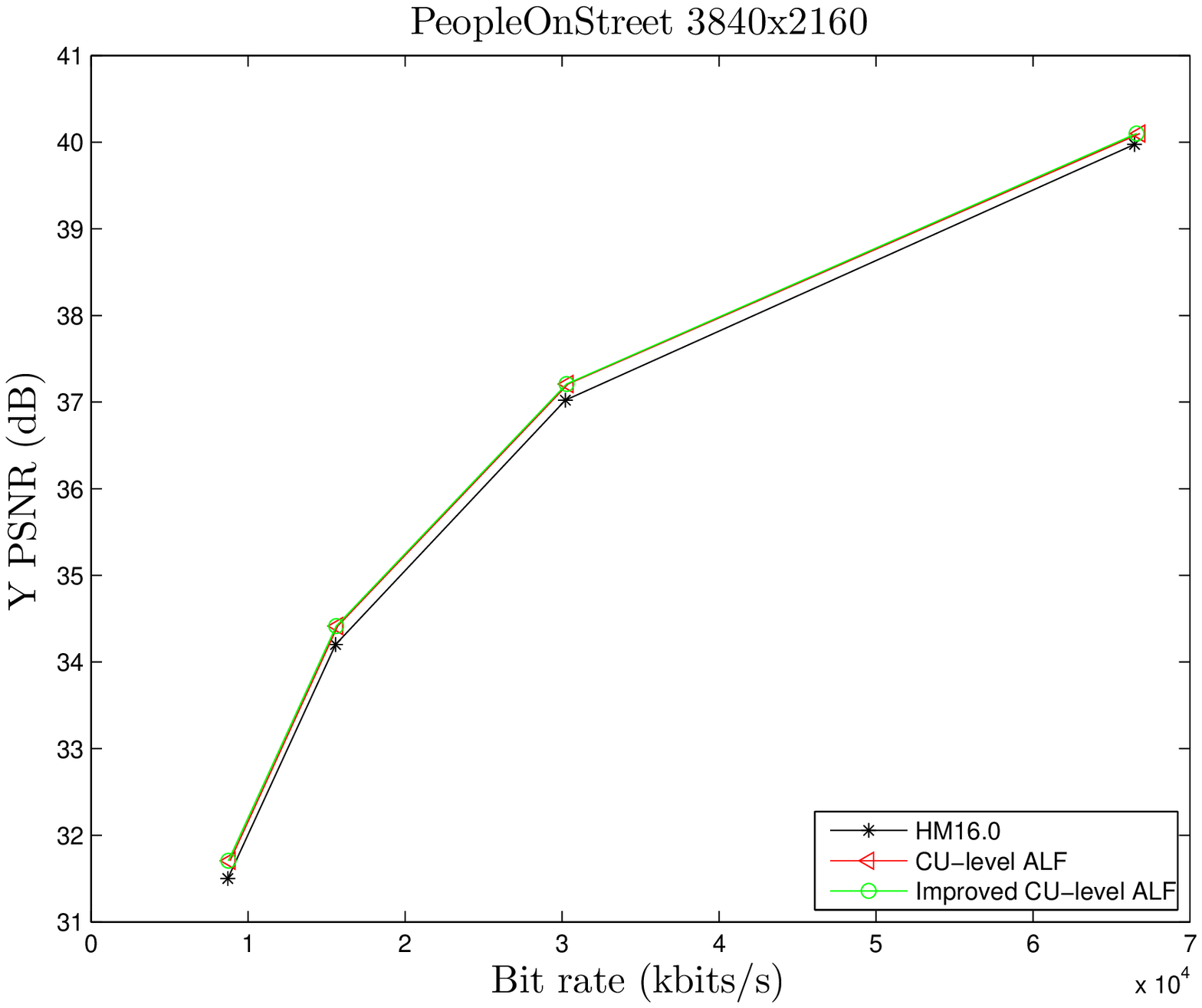}
&  \includegraphics[width=0.40 \textwidth]{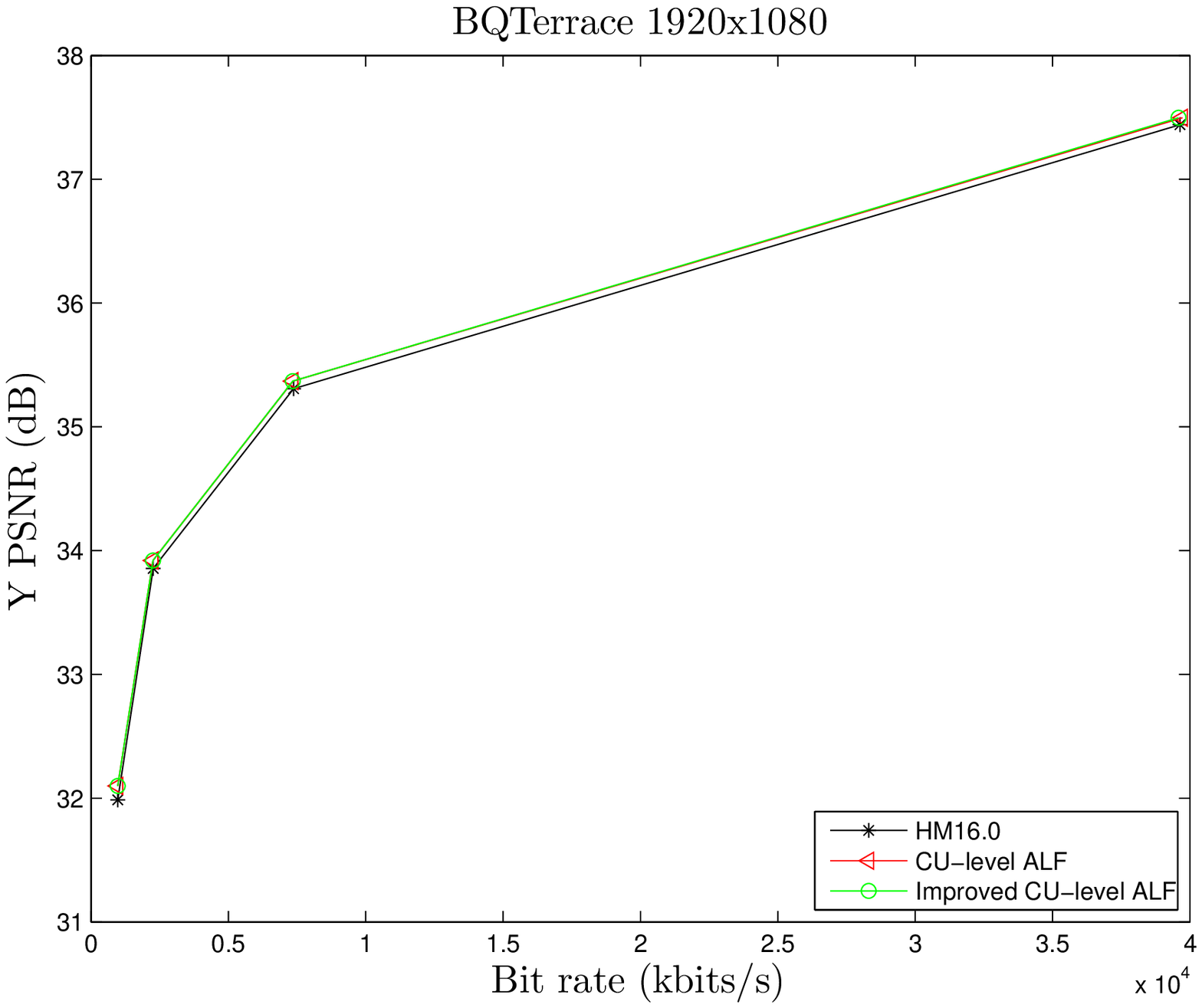}\\
 \includegraphics[width=0.40 \textwidth]{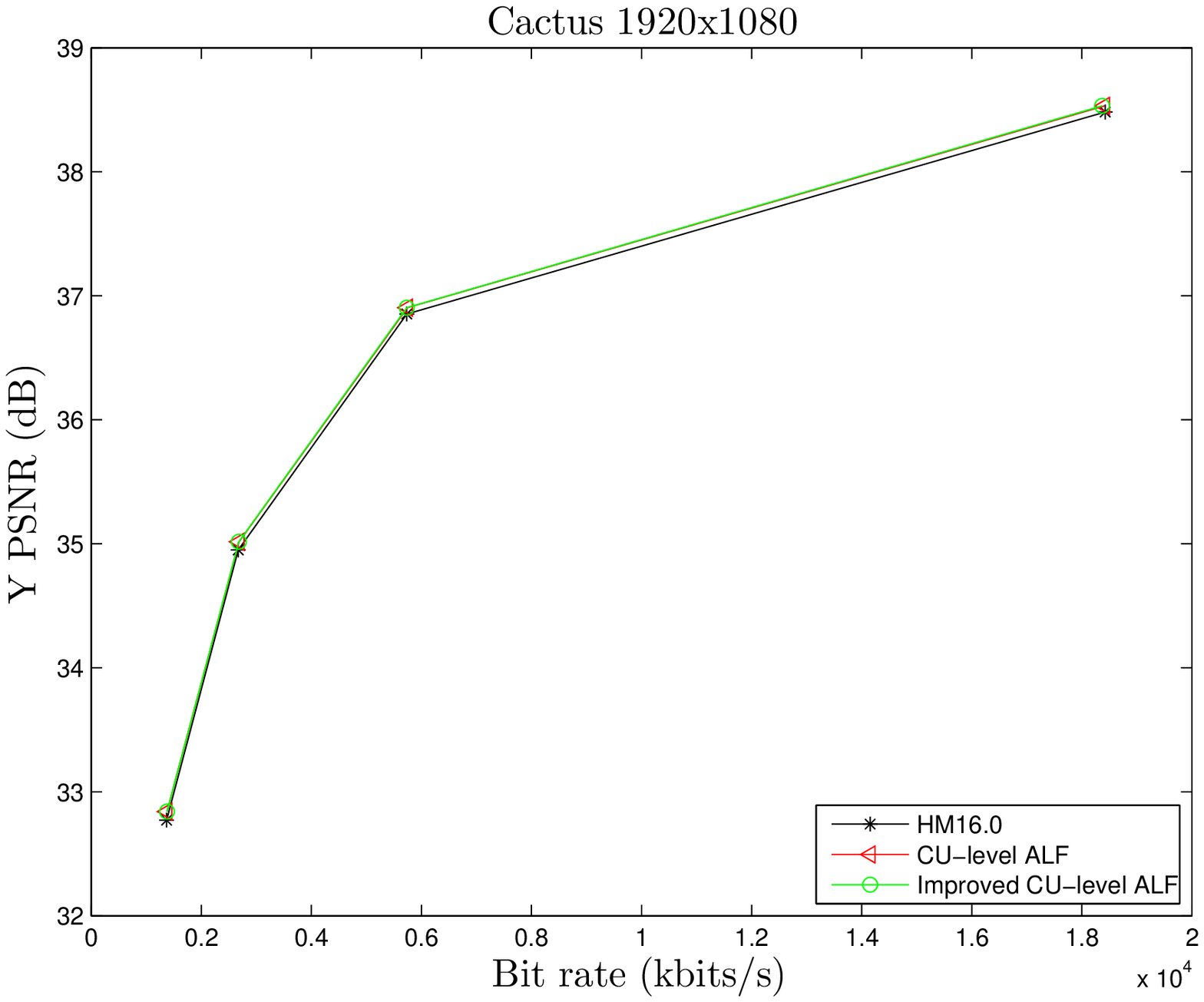}
&  \includegraphics[width=0.40 \textwidth]{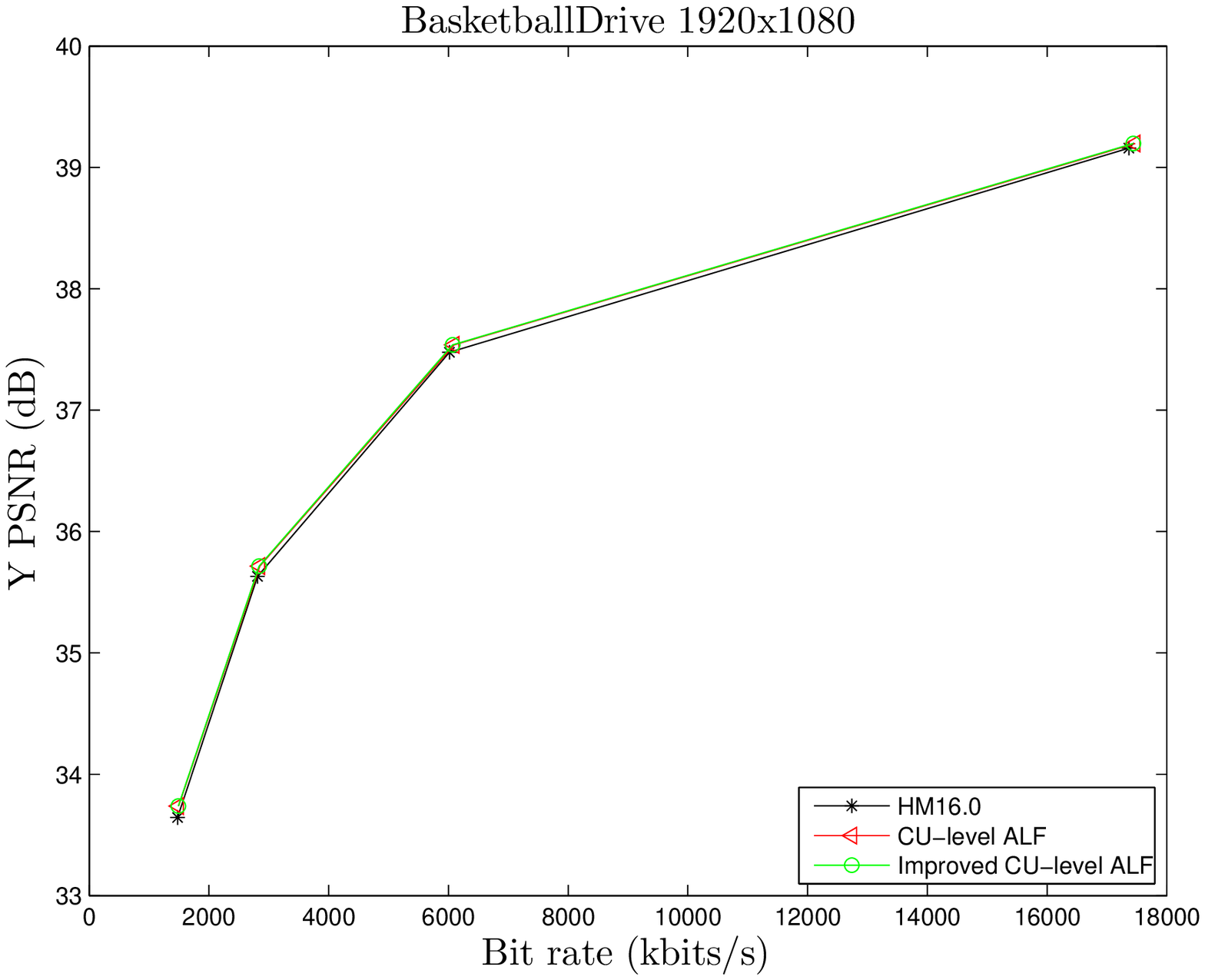}\\
\end{tabular} 
\caption{Rate-distortion comparison.}  
\label{fig:NextG:alf:rd_curves}
\end{figure*}

\subsubsection{Simulation Results}
\label{sec:NextG:alf:simulation_results}

The latest HEVC software HM16.0 is selected as the reference for comparison, where the proposed methods have been rigorously implemented and verified. In the experiments, the test settings comply with the common test conditions (CTC) \cite{JCTVC_G1200}. It should be noted that all the coding parameters are set identically for the benchmark HM16.0 and our methods. The simulation results show that the proposed method can reduce the bits of the CU level ALF.

Fig. \ref{fig:NextG:alf:rd_curves} illustrates the rate-distortion comparisons of three algorithms for the sequences \textit{PeopleOnStreet}, \textit{BQTerrace}, \textit{Cactus}, and \textit{BasketballDrive}, respectively. It can be observed that our method can achieve slightly better or comparable performance in comparison with HM16.0.

It should be noted that the increment of computation complexity for our method is trivial in comparison with the HEVC encoder. The reason is that the complexity of ALF in the encoder and decoder side has been shown to be about 5$\%$ in the standard HEVC documents, like the JCTVC-J0048  and JCTVC-J0390.   On the other hand, experimental results show that the average encoding and decoding complexities of our improved CU-level ALF method are about 0.78$\%$ and 0.90$\%$, respectively, compared to the traditional ALF method implemented on the HM16.0.

\subsection{Adaptive Sample Adaptive Offset Block Size}
\label{sec:NextG:adaptive_sao}
In the in-loop filtering process,  SAO and ALF can be employed in a cascaded way to improve the reconstructed picture quality.  The processing orders of SAO and ALF produces different coding performances. For simplicity, the processing order SAO followed by ALF is denoted as SAO-ALF, and ALF followed by SAO is denoted as ALF-SAO. The processes of SAO and ALF are dependent because they are cascaded by each other. The output of the first process will be input of the second process. However, the parameters estimation of the first process (either SAO or ALF) is independent from the parameter estimation of the second process (either ALF or SAO) at the encoder. Simulation results show that the performance of SAO-ALF is better than that of ALF-SAO.  The reason is that SAO only performs on samples which ``need'' to be compensated (i.e., it assumes that some samples are original). However, ALF aims to reduce the difference between the original and reconstructed frame. In other words, all samples are considered and filtered when ALF is on. Thus, we perform SAO before ALF in HM16.0.

However, SAO in the current HEVC standard is processed based on CTUs. Following the current design, the basic block (i.e., called SAO block) filtered by SAO will be a SCU in super-block video coding. As the size of processing unit increases, the coding efficiency of SAO will be affected. As a result, we propose to apply for SAO with a configurable size. Specifically, in the proposed method, the size of SAO block does not necessarily have to be equal to that of SCU. In other words, SAO block is independent from a SCU. The flexibility of SAO blocks can further improve the performance of the traditional SAO filter in the video coding with large resolutions.

\subsubsection{Proposed Adaptive Sample Adaptive Offset Block Size}
\label{sec:NextG:proposed_adaptive_sao}
In HEVC, the size of SAO block is same as that of CTU. The signaling of SAO parameters is at the CTU level. If the SAO filter is enabled in the current slice, the associated parameters will be placed at the beginning of the CTU bit-stream, including the merge information, type information, and offset information.  Following the current design, the SAO block will be a SCU in super-block video coding.  However, as the size of SAO processing unit increases, the effect of SAO fine granularity can be degraded. In this section, we propose to apply SAO for a SAO block that is independent of a CTU or SCU. Furthermore, an adaptive SAO block method is also proposed. As a result, this flexibility of SAO blocks can further improve the performance of the fixed SAO block method in the Ultra HD video coding.

\subsubsection{SAO With Fixed Block Size}
\label{sec:NextG:proposed_fixed_sao}
\begin{figure*}[h!]
\centering
\begin{tabular}{c c} 
 \includegraphics[width=0.35 \textwidth]{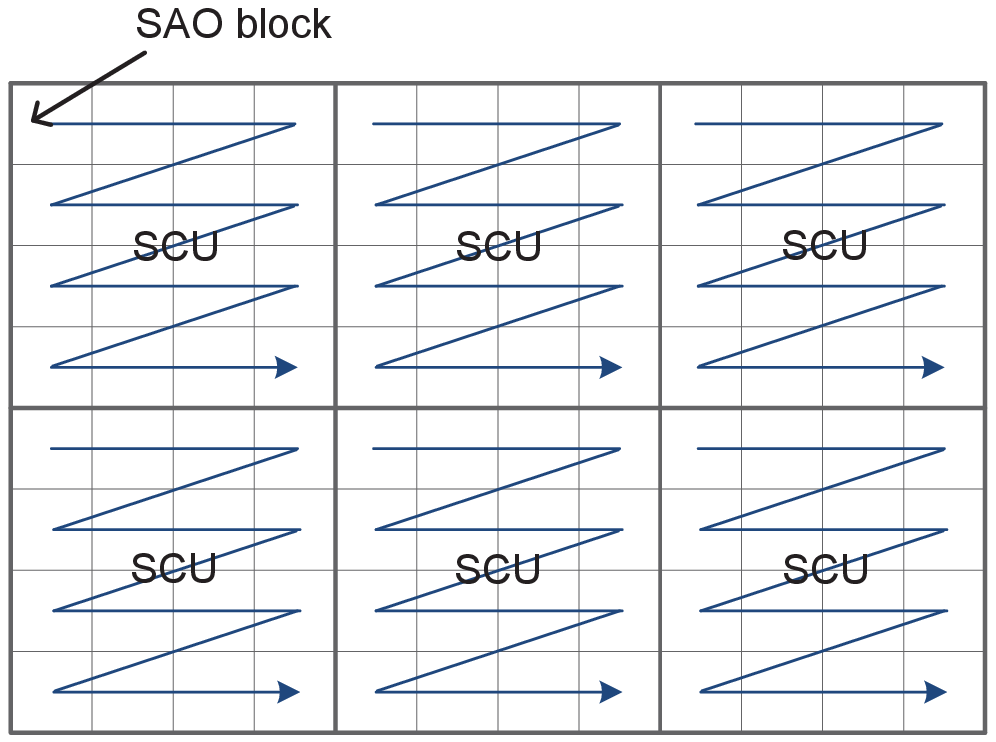}
 & \includegraphics[width=0.35 \textwidth]{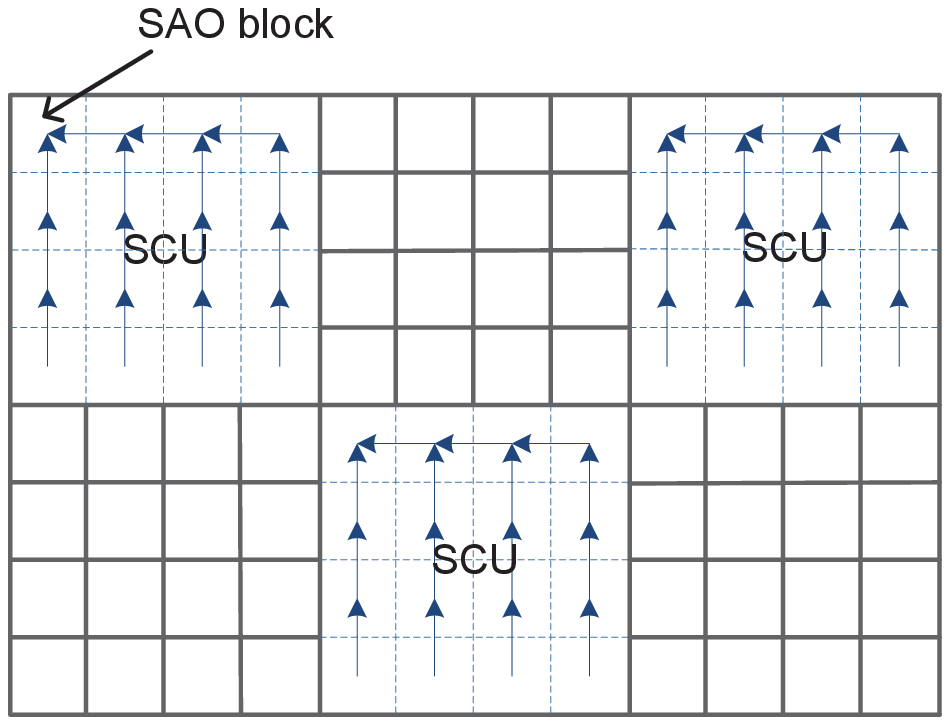} \\
 (a) & (b)
\end{tabular} 
\caption{Adaptive SAO processing unit.}  
\label{fig:NextG:sao:flowchart_fig12}
\end{figure*}

In the proposed method, each SCU or CTU consists of many SAO blocks. Commonly, all samples in a SAO block must belong to the same SCU or CTU. For each SAO block, the traditional SAO method will be applied. A typical example of the proposed SAO block structure is illustrated in Fig. \ref{fig:NextG:sao:flowchart_fig12} (a). For example, there are 6 SCUs in a frame, and for each SCU, there are total 16 SAO blocks. In a SCU, the order of SAO process follows the raster scanning order as shown in Fig. \ref{fig:NextG:sao:flowchart_fig12} (a).

\subsubsection{SAO With Adaptive Block Size}
\label{sec:NextG:proposed_adaptive_sao}

In this section, we propose an improved SAO method, called adaptive SAO with variable block size. The main idea is that the basic block sizes processed by SAO can be adaptively determined based on the overall rate distortion cost for the current SCU, where two difference SAO block sizes can be supported, and one is the SCU while the other is the predefined SAO block size as discussed before.

The adaptive SAO method takes the SAO block as the basic storage unit, where the size of SAO block is no larger than that of SCU. Fig. \ref{fig:NextG:sao:flowchart_fig12} (b) shows a typical example of the proposed SAO method. The super-block consisted of dashed line blocks indicates that the SAO size is equal to the SCU. All parameters for the current SAO processing unit is stored in the top-left SAO block, while the rest dashed SAO blocks will take the merge mode whose direction is represented by arrow as shown in Fig. \ref{fig:NextG:sao:flowchart_fig12} (b).   The super-block consisted of solid line block indicates that the SAO size is equal to a predefined SAO block (smaller than the size of SCU). For these solid line blocks, each SAO block requires to choose and send its own SAO encoding mode (i.e., OFF, NEW or MERGE) based on the rate distortion cost minimization.

There is a SCU level SAO split flag that can be used to indicate the size of the current SAO processing unit in the bitstream. The SAO size flag is used to save the merge information in the dashed blocks, since the decoder can derive the associated merge information according to the status of this flag.

\begin{table}[!h]
\centering
\caption{Comparisons of SAO methods  under the low-delay configuration.} 
\label{tab:NextG:sao:result_Y_ld}
\begin{lrbox}{\tablebox}
\begin{tabular}{|l|ccc|ccc|}
\hline
		&\multicolumn{3}{|c|}{Fixed SAO block}     &\multicolumn{3}{|c|}{Adaptive SAO block} \\
\cline{2-7}
Sequence				&BD-Bitrate &BD-Bitrate & BD-Bitrate  &BD-Bitrate &BD-Bitrate & BD-Bitrate\\
                        &(Y)		&(U)		&(V)	      &(Y)		  &(U)		  &(V)\\
\hline                      
Traffic	&-0.6$\%$	&-1.1$\%$	&-1.3$\%$	&-0.3$\%$	&-1.4$\%$	&-1.2$\%$\\
PeopleOnStreet	&-0.2$\%$	&-0.7$\%$	&-1.0$\%$	&-0.1$\%$	&-1.3$\%$	&-2.4$\%$\\
SDR$\_$candlelight	&-2.9$\%$	&-4.1$\%$	&-3.6$\%$	&-4.5$\%$	&-14.3$\%$	&-14.9$\%$\\
SDR$\_$rainfruits	&-1.0$\%$	&-2.2$\%$	&-2.0$\%$	&-2.1$\%$	&-11.2$\%$	&-8.3$\%$\\
Cactus	&-0.5$\%$	&-1.4$\%$	&-1.8$\%$	&0.9$\%$	&-0.3$\%$	&0.4$\%$\\
BQTerrace	&-0.7$\%$	&-1.5$\%$	&-2.8$\%$	&0.1$\%$	&-4.0$\%$	&-7.8$\%$\\
BasketballDrive	&-0.6$\%$	&-1.4$\%$	&-1.4$\%$	&-0.7$\%$	&-3.7$\%$	&-2.9$\%$\\
Candlelight10	&-2.8$\%$	&-5.4$\%$	&-2.2$\%$	&-4.6$\%$	&-15.8$\%$	&-14.0$\%$\\
Rainfruits10	&-1.1$\%$	&-2.1$\%$	&-2.3$\%$	&-2.2$\%$	&-10.9$\%$	&-8.3$\%$\\
Birthday10	&-2.1$\%$	&-2.3$\%$	&-1.8$\%$	&-2.8$\%$	&-4.0$\%$	&-3.0$\%$\\
Market10	&-0.7$\%$	&-1.9$\%$	&-3.2$\%$	&0.5$\%$	&-8.4$\%$	&-8.4$\%$\\ \hline
Average	&-1.2$\%$	&-2.2$\%$	&-2.1$\%$	&-1.4$\%$	&-6.8$\%$	&-6.4$\%$\\

\hline
\end{tabular}
\end{lrbox}
\scalebox{0.70}{\usebox{\tablebox}}
\end{table}

\begin{table}[!h]
\centering
\caption{Comparisons of SAO methods under the random-access configuration.} 
\label{tab:NextG:sao:result_Y_ra}
\begin{lrbox}{\tablebox}
\begin{tabular}{|l|ccc|ccc|}
\hline
		&\multicolumn{3}{|c|}{Fixed SAO block}     &\multicolumn{3}{|c|}{Adaptive SAO block} \\
\cline{2-7}
Sequence				&BD-Bitrate &BD-Bitrate & BD-Bitrate  &BD-Bitrate &BD-Bitrate & BD-Bitrate\\
                        &(Y)		&(U)		&(V)	      &(Y)		  &(U)		  &(V)\\
\hline                      
Traffic	&-0.8$\%$	&-1.2$\%$	&-1.0$\%$	&-0.9$\%$	&-1.4$\%$	&-0.8$\%$\\
PeopleOnStreet	&-0.1$\%$	&-0.9$\%$	&-0.8$\%$	&-0.3$\%$	&-1.9$\%$	&-2.5$\%$\\
SDR$\_$candlelight	&-3.1$\%$	&-5.8$\%$	&-5.5$\%$	&-3.6$\%$	&-14.2$\%$	&-13.2$\%$\\
SDR$\_$rainfruits	&-1.0$\%$	&-1.5$\%$	&-1.3$\%$	&-1.9$\%$	&-5.2$\%$	&-3.8$\%$\\
Cactus	&-0.3$\%$	&-1.1$\%$	&-1.0$\%$	&0.0$\%$	&-3.4$\%$	&-1.4$\%$\\
BQTerrace	&-0.8$\%$	&-1.6$\%$	&-1.1$\%$	&-0.6$\%$	&-2.4$\%$	&-2.4$\%$\\
BasketballDrive	&-0.9$\%$	&-1.1$\%$	&-1.1$\%$	&-1.0$\%$	&-1.5$\%$	&-1.6$\%$\\
Candlelight10	&-3.1$\%$	&-5.5$\%$	&-5.9$\%$	&-3.6$\%$	&-13.9$\%$	&-13.0$\%$\\
Rainfruits10	&-1.0$\%$	&-1.9$\%$	&-1.7$\%$	&-1.9$\%$	&-5.5$\%$	&-4.0$\%$\\
Birthday10	&-2.8$\%$	&-2.2$\%$	&-2.2$\%$	&-3.0$\%$	&-2.6$\%$	&-2.7$\%$\\
Market10	&-0.7$\%$	&-2.5$\%$	&-2.6$\%$	&-0.2$\%$	&-4.3$\%$	&-3.5$\%$\\  \hline
Average	&-1.3$\%$	&-2.3$\%$	&-2.2$\%$	&-1.5$\%$	&-5.1$\%$	&-4.4$\%$\\

\hline
\end{tabular}
\end{lrbox}
\scalebox{0.70}{\usebox{\tablebox}}
\end{table}

\subsubsection{Simulation Results}
\label{sec:NextG:sao:simulation_results}

The proposed adaptive SAO algorithm has been implemented in HM16.0 which is also used as a benchmark to evaluate our method. Tables \ref{tab:NextG:sao:result_Y_ld} and \ref{tab:NextG:sao:result_Y_ra} summarize the simulation results of low-delay and random-access, respectively. As can be seen from the above two tables,  significant coding gains of the U-component and V-component are achieved. Specifically, average bit reductions of 4.6$\%$ and 4.3$\%$ are obtained in the low-delay configuration, and average bit reductions of 2.8$\%$ and 2.2$\%$ are obtained in the random-access configuration.

The computational complexity of our method is also evaluated. As can be seen in Section \ref{sec:NextG:adaptive_sao}, the additional operation of our method is the summation of the rate-distortion cost of SAO blocks. Since the implementation of SAO is very efficient in HM16.0, the additional complexity of our method is acceptable based on its coding performance. Simulation results show that the average complexity of our method is about 1.2$\%$ compared to that of HM16.0 at the encoder.

\section{Summary}
\label{Sec:summary}

The growing needs for high-quality video applications have resulted in a lot of studies and developments in video signal coding. This chapter presents some advanced techniques in enhancing the rate-distortion performance of the block-based hybrid video coding systems.  Additionally, as can be seen from the developments of H.264/AVC and HEVC, most of the current coding tools, such as prediction, transformation and entropy coding, have less room to improve in the compression performance. On the other hand, loop filer in the modern video standards shows the promising results. Thus, we believe that loop filter can be the candidate in contributing to higher video compression for the next-generation video coding. Specifically, improvements on ALF and SAO are also introduced, and the simulation results show that the proposed methods outperform the existing method, which offer new degrees of freedom to improve the overall rate-distortion performance. As a result, they can be the candidate coding tools for the next-generation video codec.

The main work introduced in this chapter is reorganized from my PhD thesis ``Adaptive Coding and Rate Control of Video Signals'', the Chinese University of Hong Kong, Nov. 2015. Note that methods presented in Section \ref{Sec:STDS:future} were done while the author was a visiting scholar at the Innovation Laboratory of InterDigital Communications Corporation (IDCC), San Diego, USA.

\bibliography{intech-book} 
\end{document}